\documentclass{article}
\usepackage[utf8]{inputenc}
\usepackage{natbib}
\usepackage{authblk}
\usepackage{graphicx}
\usepackage{url}
\usepackage{amsmath}
\usepackage{hyperref}

\def\beq{\begin{equation}}
\def\eeq{\end{equation}}
\def\bea{\begin{eqnarray}}
\def\eea{\end{eqnarray}}

\title{Electric charge of black holes: Is it really always negligible?\\ \vspace{1cm} Submitted to the Observatory}
\author[1]{\textbf{Michal Zaja\v{c}ek}}
\author[2]{\textbf{Arman Tursunov}}
\affil[1]{Center for Theoretical Physics, Polish Academy of Sciences, Al. Lotnikow 32/46, 02-668 Warsaw, Poland;\\email: \href{mailto: zajacek@ph1.uni-koeln.de}{\textbf{zajacek@ph1.uni-koeln.de}}}
\affil[2]{Institute of Physics and Research Centre of Theoretical Physics and Astrophysics, Faculty of Philosophy and Science, Silesian University in Opava, Bezru\v{c}ovo n\'{a}m.13, CZ-74601 Opava, Czech Republic; \\email: \href{mailto: arman.tursunov@fpf.slu.cz}{\textbf{arman.tursunov@fpf.slu.cz}}}

\begin{document}
\maketitle

\begin{abstract}
 We discuss the problem of the third black hole parameter, an electric charge. While the mass and the spin of black holes are frequently considered in the majority of publications, the charge is often neglected and implicitly set identically to zero. However, both classical and relativistic processes can lead to a small non-zero charge of black holes. When dealing with neutral particles and photons, zero charge is a good approximation. On the other hand, even a small charge can significantly influence the motion of charged particles, in particular cosmic rays, in the vicinity of black holes. Therefore, we stress that more attention should be paid to the problem of a black-hole charge and hence, it should not be neglected \textit{a priori}, as it is done in most astrophysical studies nowadays. The paper looks at the problem of the black-hole charge mainly from the astrophysical point of view, which is complemented by a few historical as well as philosophical notes when relevant. In particular, we show that a cosmic ray or in general elementary charged particles passing a non-neutral black hole can experience an electromagnetic force as much as sixteen times the gravitational force for the mass of the Galactic centre black hole and its charge being seventeen orders of magnitude less than the extremal value (calculated for a proton). Furthermore, a Kerr-Newman rotating black hole with the maximum likely charge of 1 Coulomb per solar mass can have the position of its innermost stable circular orbit (ISCO) moved by both rotation and charge in ways that can enhance or partly cancel each other, putting the ISCO not far from the gravitational radius or out at more than 6 gravitational radii. An interpretation of X-ray radiation from near the ISCO of a black hole in X-ray binaries is then no longer unique.     
\end{abstract}

\newpage
  Although initially believed to be out of the real Universe like ``unicorns and gargoyles'' \citep{thorne1994black}, black holes are not \textit{corps obscurs} anymore. Instead, over the last decades, they have been fully accepted as real astrophysical entities. They form an integral part of the stellar evolution and  even more importantly, ``feeding'' massive black holes and the associated feedback appear to be crucial to fully account for the galaxy evolution.
  
  Experimental means to study black holes are now richer than ever. Since the end of the 1960s, their footprints have been successfully studied via multi-wavelength electromagnetic-based observations. At the centenary of the theory of general relativity, a new channel was opened up thanks to the first detection of gravitational waves that resulted from the merger of two stellar black holes \citep{2016PhRvL.116f1102A}. A recent detection of a high-energy neutrino in combination with the quasi-simultaneous $\gamma$-ray counterpart has enabled astrophysicists to pinpoint its origin to a supermassive black hole with the relativistic jet directed almost exactly towards us -- BL Lac object TXS 0506$+$056 \citep{147,2019A&A...630A.103B}. This has opened a new era of the so-called `multi-messenger astronomy’  -- the term implying observations of four disparate cosmic ``messengers”: electromagnetic radiation, gravitational waves, neutrinos, and cosmic rays.

According to the general relativity, any information about the black hole matter is hidden inside its event horizon, being inaccessible to external observers, which is referred to as the ``no-hair'' theorem or rather conjecture \citep{1973grav.book.....M}. This makes it possible to describe any astrophysical black hole by just three classical, externally observable parameters: its mass, its spin (angular momentum), and electric charge (in case we don't consider the speculative magnetic monopole at this point). 
  One of the main motivations behind the exciting, and often time-demanding multi-messenger experiments is to determine the black hole mass and its spin. The third parameter, electric charge, is usually neglected and basically set equal to zero. This assumption is often backed up by arguing that the presence of plasmas around astrophysical black holes leads to prompt discharging. The negligence of charge has also affected the theoretical investigation of the particle motion in the vicinity of Reissner-Nordstr\"om black hole \citep{1916AnP...355..106R,1918KNAB...20.1076N}. Studies of the motion of charged test particles such as the paper by Pugliese et al. \citep{2011PhRvD..83j4052P} could in principle have appeared decades earlier in case the astrophysical motivation for charge had been bigger.    
  
However, is the black-hole charge always exactly zero? Hasn't it been neglected too often just to simplify calculations? And if there is any charge, can it lead to some observable effects?
  
  First of all, how could black holes get charged? It was already pointed out by Arthur S. Eddington \citep{1926ics..book.....E} that stars should bear a small positive charge to prevent electrons and protons from further separation in the stellar atmosphere due to the mass difference by a factor of nearly $2\,000$. To get an estimate of this charge, we consider the combined conservative gravitational and electric field around a black hole, $\phi(\mathbf{r})=\phi_{\rm G}(\mathbf{r})+\phi_{\rm E}(\mathbf{r})$, where the corresponding force is $\mathbf{F}=-\nabla{\phi(\mathbf{r})}$. The distribution function $f(\mathbf{r},\mathbf{w})$ for the Maxwell-Botzmann equilibrium distribution (MW) in the external conservative field can be expressed as,
\begin{equation}
  f(\mathbf{r},\mathbf{w})=n_0(\mathbf{r})\exp{\left[-\frac{\phi(\mathbf{r})}{k_{\rm B}T}\right]}f(\mathbf{w})\,,
  \label{eq_maxwell_boltzmann_external}
\end{equation}  
where $n_0(\mathbf{r})$ is the density distribution in the absence of the external field and $f(\mathbf{w})$ is the velocity-dependent part of the MW distribution. The density distribution of a particle species is simply given by,

\begin{equation}
  n_{\rm par}(\mathbf{r})=n_0(\mathbf{r})\exp{\left[-\frac{\phi(\mathbf{r})}{k_{\rm B}T}\right]}\,.
  \label{eq_density_distribution}
\end{equation}
Since we expect that the density distribution of electrons and protons is comparable at any distance to ensure quasineutrality around astrophysical bodies including black holes, $n_{\rm p} \approx n_{\rm e}$, it follows from Eq.~\eqref{eq_density_distribution} that the potential value for protons and electrons should also be approximately the same, $\phi_{\rm p}\approx  \phi_{\rm e}$. From the potential equality, we obtain a value of the equilibrium charge $Q_{\rm eq}$ and the charge to mass ratio can be expressed in terms of fundamental constants,
\begin{equation}
\frac{Q_{\rm eq}}{M_{\bullet}}=\frac{2\pi \epsilon_0 G(m_{\rm p}-m_{\rm e})}{e}\approx 76.9\, C\,M_{\odot}^{-1}\,.
\label{eq_charge_separation}
\end{equation}

   This was generalized by John Bally and ``Ted'' Harrison \citep{1978ApJ...220..743B} at the end of 1970s, who derived that any macroscopic body in the Universe -- stars, galaxies, and therefore also black holes -- are positively charged with the charge-to-mass ratio of approximately 100 Coulombs per Solar mass. In this ``electrically polarized Universe'', the positive charge of galaxies is compensated by a negatively charged, freely expanding intergalactic medium.
  
  Another mechanism that supports the existence of charged black holes is purely relativistic. In the same way as space and time are fundamentally the same, being just the different components of the four-dimensional space-time coordinates, electric and magnetic fields are also just the different components of the antisymmetric, rank 2 tensor of an electromagnetic field, $F_{\mu\nu}\equiv \partial_{\mu}A_{\nu}-\partial_{\nu}A_{\mu}$, where $A_{\nu}$ is the electromagnetic potential. It appears that a rotating black hole immersed within the external magnetic field (produced by e.g. a dynamo of plasma around a black hole) in fact induces an electric field due to the twisting of magnetic field lines. This was shown in 1974 by Robert M. Wald \citep{1974PhRvD..10.1680W}. A convincing number of evidence that magnetic fields are indeed present in the vicinity of astrophysical black holes alongside the fact that any  black hole is generally rotating therefore imply that a non-zero charge of a black hole is quite plausible.
A value of such an induced charge is proportional to both the strength of the magnetic field and the spin of a black hole, which is also recovered and applied in the recent studies focused on the Wald mechanism \citep{2018MNRAS.480.4408Z,2018PhRvD..98l3002L}. The rotation of a black hole in the ordered external magnetic field leads to the Faraday induction, where the time-component of the electromagnetic potential represents the induced electric field. The potential difference between the black hole horizon and infinity is the following,

\beq
\Delta \phi = \phi_{\rm H} - \phi_{\infty} = \frac{Q_{\bullet} - 2 a_{\bullet} M_{\bullet} B_{\rm ext}}{2 M_{\bullet}}\,,
 \label{eq_potential_difference}
\eeq 
where $Q_{\bullet}$, $a_{\bullet}$, and $M_{\bullet}$ are the charge, dimensionless spin, and the mass of the black hole, respectively, and $B_{\rm ext}$ is a magnitude of the external homogeneous magnetic field. The potential difference expressed by Eq.~\eqref{eq_potential_difference} leads to the selective accretion of charges or in other words, the charging of the black hole. The charging stops when the potential difference is zero, which occurs for the maximum net charge of $Q_{\bullet}=2a_{\bullet}M_{\bullet}B_{\rm ext}$. Considering the supermassive black hole at the Galactic centre  with the mass of $M_{\bullet}=4.14\times 10^6\,M_{\odot}$ immersed in the poloidal magnetic field of $10\,{\rm G}$ \citep{2018A&A...615L..15G}, the induced charge is limited by the maximum spin of $a_{\bullet}\leq M_{\bullet}$,

\beq 
Q_{\bullet \rm ind}^{\rm max} \leq 2.5 \times 10^{15} \left( \frac{M_{\bullet}}{4.14 \times 10^6 M_{\odot}} \right)^2  \left( \frac{B_{\rm ext}}{10 \rm G} \right)  ~\rm C\,.
\label{eq_max_induced_charge}
\eeq

 A sign of the Wald charge depends on the orientation of magnetic field lines with respect to the black hole spin. If the magnetic field is directed alongside the rotation vector of a black hole, then the charge is positive. This implies that, as for stars and galaxies as a whole, the charge of astrophysical black holes tends to be positive since a certain degree of an alignment between the accretion flow angular momentum and the black hole spin is expected on a sufficiently long time-scale.
  
  Even if black holes are charged with the charge-to-mass ratio of about 100 Coulombs per Solar mass, the motion of neutral bodies as well as photons in their vicinity is not significantly affected. Therefore the usual assumption of zero charge appears to be fair enough. Just to get some specific estimates, for the Galactic center black hole of 4 million Solar masses \citep{2017FoPh...47..553E}, we would expect the charge of about $10^8$ Coulombs based on the electron-proton separation \citep{2018MNRAS.480.4408Z}, see Eq.~\eqref{eq_charge_separation}. The limiting, extremal charge of such a massive black hole, which would noticeably affect the space-time metric, is nighteen orders of magnitude larger, which follows from the general Kerr-Newman black hole solution,
  
\begin{equation}
  Q_{\rm max}^{{\rm rot}}=2M_{\bullet}\sqrt{\pi \epsilon_0 G (1-a_{\bullet}^2)}\,,
  \label{eq_max_gen_nodim}
\end{equation}
which for the non-rotating, Reissner-Nordstr\"om case ($a_{\bullet}=0$) may simply be evaluated as
\begin{equation}
 Q_{\rm max}^{{\rm norot}}=2\sqrt{\pi \epsilon_0 G}M_{\bullet}=6.86 \times 10^{26}\, \left(\frac{M_{\bullet}}{4\times 10^6\,M_{\odot}} \right)\,C\,.
 \label{eq_max_charge}
\end{equation}

The value of an induced charge due to the black hole rotation in the surrounding magnetic field of about 10 Gauss, as inferred from the flaring activity of the Galactic center black hole \citep{2017FoPh...47..553E}, has an upper limit of about $10^{15}$ Coulombs, see Eq.~\eqref{eq_max_induced_charge}, which is still not large enough to have a noticeable impact on the space-time geometry \citep{2018MNRAS.480.4408Z}.  Hence, so far the assumption of a zero charge seems to be on the safe side.
  
  However, this assumption is only valid when one considers the motion of neutral matter including photons. The dynamics of charged particles, such as electrons and protons, can be profoundly affected in case the black hole possesses even very small charge, as e.g. the likely value of $10^8$ Coulombs for the Galactic center black hole. So what is then the consequence of this charge?  In contrast to a magnetic field, an electric field can do work and consequently, it can support the acceleration of charged particles to very large, relativistic velocities. In particular, the charge of a rotating black hole generates the electromotive force between the pole of the black hole and its equator. A subsequent discharge of the black hole due to the accretion of oppositely charged matter slows down the black-hole rotation. This process became known as the Blandford-Znajek mechanism \citep{1977MNRAS.179..433B} which is generally considered to be one of the main processes for the generation of relativistic jets. This is how the rotational energy of black holes can be extracted out and the charge is the driving engine in this case.
  
    Three out of the four available cosmic `messengers' -- photons, gravitational waves, and neutrinos -- are neutral particles that are naturally not affected by electromagnetic fields. This makes it possible to trace their origin even if the source is located very far from the Earth. Hence, the assumption of zero charge seems reasonable in the construction of theories of their formation.
  This remains true until the fourth messenger, the cosmic rays, comes into play. These are charged particles that are detected with energies unreachable by current particle accelerators. Recent discoveries \citep{147} have traced their origin to supermassive black holes. However, the question why the energies are so high still remains mysterious.
  So not only are ultra-high energy cosmic rays the most energetic particles, they are also the most baffling ones. The non-zero charge of a black hole would lead to an inevitable electromagnetic interaction of the black hole with cosmic-ray particles in their source region. This interaction can be much stronger than the standard gravitational one, since the ratio of the electrostatic and gravitational forces acting on a charged particle of charge $q_{\rm par}$ and mass $m_{\rm par}$ is larger than unity for even small charges of the massive black hole $Q_{\bullet}\ll Q_{\rm max}^{\rm rot}$,
\begin{equation}
  \frac{F_{\rm elstat}}{F_{\rm grav}}=\frac{1}{4\pi \epsilon_0 G}\left(\frac{Q_{\bullet}}{M_{\bullet}}\right)\left(\frac{q_{\rm par}}{m_{\rm par}} \right)\simeq 16 \left(\frac{Q_{\bullet}}{10^{10}\,C}\right) \left(\frac{M_{\bullet}}{4\times 10^6\,M_{\odot}}\right)^{-1}\,,  
\end{equation}    
where the last equality was evaluted for a proton in the vicinity of the black hole of $4\times 10^6$ Solar masses.
   For the discussion of a potential effect of the black hole charge on energy extraction mechanisms, see a review article by Zaja\v{c}ek et al. \citep{2018arXiv181203574Z}. So, maybe the key for the demystification of ultra-high energy cosmic rays lies in the neglected charge?
   
   \begin{figure}[h!]
      \centering
      \includegraphics[width=\textwidth]{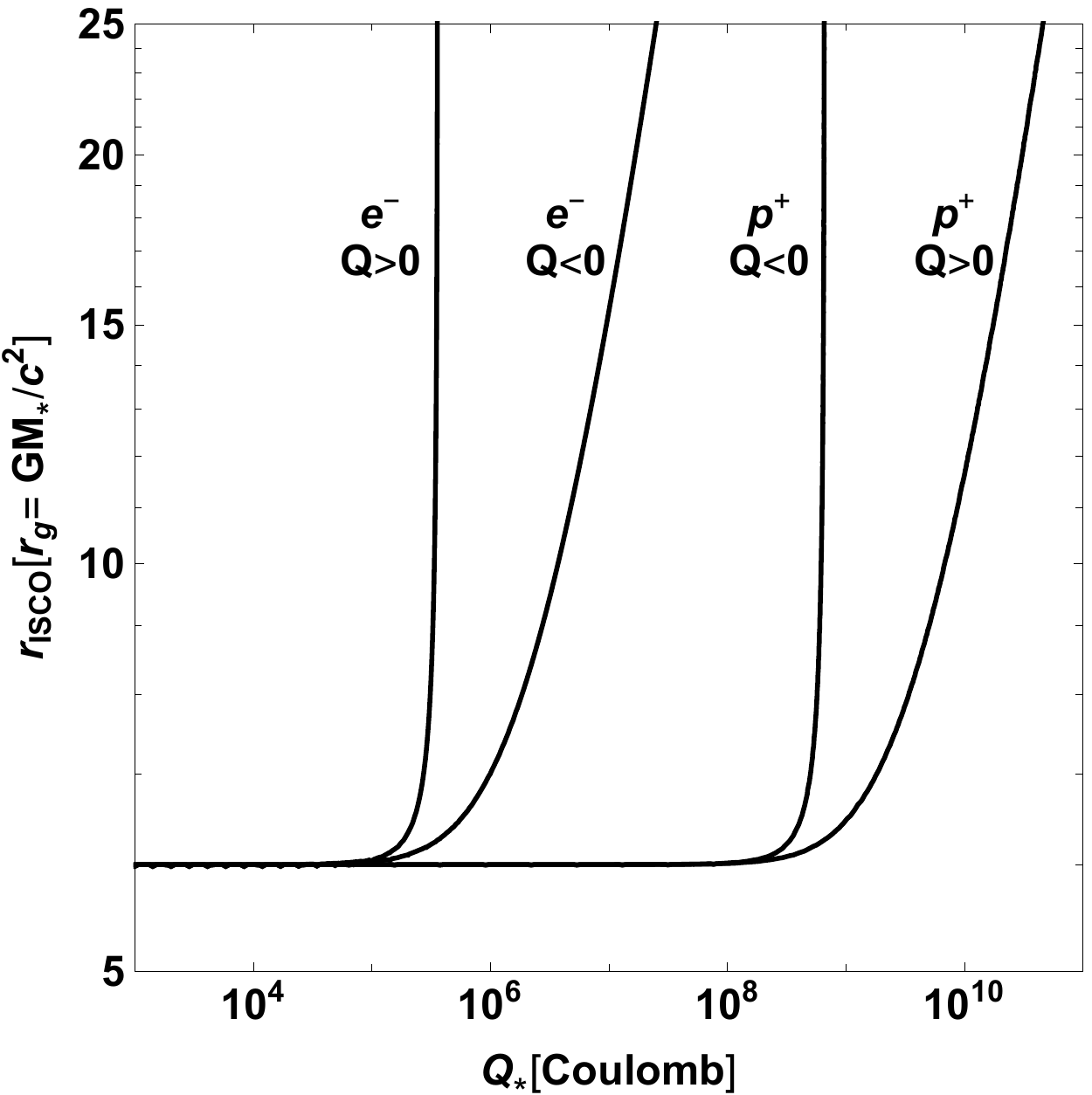}
      \caption{The ISCO location (expressed in gravitational radii, $r_{\rm g}=GM_{\bullet}/c^2$) for a non-rotating black hole as a function of a small electric charge (expressed in Coulombs) for the black hole at the Galactic centre with the mass of $\sim 4$ million Solar masses. The ISCO location for a non-rotating, uncharged case is at $r_{\rm g}=6GM_{\bullet}/c^2$.}
      \label{fig_ISCO_charge}
   \end{figure}
  
  One of the crucial parameters in the accretion theory is the location of the innermost stable circular orbit (ISCO) of orbiting matter. For neutral non-rotating black holes, the ISCO is located at the distance of six gravitational radii from the singularity, $r_{\rm ISCO}=6GM_{\bullet}/c^2$. For rotating black holes, the ISCO shifts closer, up to the event horizon at $r_{\rm ISCO}=GM_{\bullet}/c^2$, for prograde rotation, i.e. for a black hole rotating in the same sense as the orbiting matter. In contrast, for black holes that rotate in a retrograde sense, the ISCO shifts further away up to nine gravitational radii, $r_{\rm ISCO}=9GM_{\bullet}/c^2$. In a similar way as the spin, electric charge by itself can also shift the ISCO, most profoundly for charged particles \citep{2011PhRvD..83j4052P}. The electric charge shifts the ISCO away from the black hole for like and opposite charges, respectively. Hence, it mimicks the retrograde black hole spin. This is demostrated for a non-rotating black hole with a small electric charge in Figure~\ref{fig_ISCO_charge} for all plausible cases that can occur: a positively charged black hole with an electron or a proton around it and a similar set-up for a negatively charged black hole. As can be seen from Figure~\ref{fig_ISCO_charge}, four different situations can arise according to the charges of the black hole and a particle and its mass. The ISCO at $9GM_{\bullet}/c^2$ can be reached for the small black hole charge of only $Q_{\bullet}=3.3 \times 10^5 {\rm C}$ and $Q_{\bullet}=2.8 \times 10^6\,{\rm C}$ for electrons for the positively and the negatively charged black hole, respectively. Similarly, for protons, the ISCO shifts to $9GM_{\bullet}/c^2$ for the negative black hole charge of $Q_{\bullet}=6.0 \times 10^8\,{\rm C}$ and the positive charge of $Q_{\bullet}=5.07 \times 10^9\,{\rm C}$. Furthermore, the circumnuclear magnetic field can lead to both an inward and outward shift of the ISCO with respect to the black hole \citep{2016PhRvD..93h4012T}. As a consequence, it leads to a certain degree of \textit{underdetermination} of what can cause the actual shift of the ISCO. In other words, the circumnuclear magnetic field and a small charge can mimic the black hole spin and it can become quite intricate for observers to decide who the real ``culprit'' shifting the ISCO is. An interpretation of electromagnetic radiation from close to the ISCO of black holes, for instance X-ray light curves in X-ray binaries or the Galactic centre black hole, is thus no longer unique.      
  
  In conclusion, the black-hole charge may not be just a purely ``academic'' parameter, with no relevance to observations. More attention should be paid to its potential effect on the dynamics of charged particles in the direct grasp of black holes, and thus it should not be neglected \textit{a priori}, as it is routinely done in most astrophysical studies. 
   
\section*{Acknowledgements}
We thank the referee, Virginia Trimble, for useful comments that improved the draft. We are also grateful to Martin Kolo\v{s} for many discussions about the problem and the help with the manuscript preparation. Michal Zaja\v{c}ek acknowledges the financial support from the National Science Centre, Poland, grant No. 2017/26/A/ST9/00756 (Maestro 9).

\newpage  
\bibliographystyle{apj}
\bibliography{charge}

\end{document}